\documentclass[preprint,12pt]{aastex}

\usepackage{graphicx}
\usepackage{amsmath}
\usepackage{amssymb}

\begin{document}

\title{GRB 990123 revisited: Further Evidence for a Reverse Shock}

\author{Ehud Nakar \altaffilmark{1,2} and Tsvi Piran \altaffilmark{1}}

\altaffiltext{1}{Racah Institute for Physics, The Hebrew
University, Jerusalem 91904, Israel}

\altaffiltext{2}{ Institut d`Astrophysique de Paris, 75014 Paris,
France}

\begin{abstract}
Recently we have presented a new theoretical analysis of the
reverse shock emission. We use this analysis here  to revisit GRB
990123. We find new and compelling evidences  that the optical
flash and the radio flare of GRB 990123 resulted from a reverse
shock. This suggests that a significant fraction of the energy of
the relativistic ejecta must have been carried by baryons. It also
suggests that the external medium is an ISM and that in this burst
the reverse shock emission dominated at early time over other
possible processes. We use the early optical emission to constrain
the physical parameters of the original ejecta and the microscopic
parameters in the emitting reverse shocked region.
\end{abstract}
\maketitle

\section{Introduction}

The 9th magnitude optical flash of GRB 990123 was one of the most
exciting discoveries associated with Gamma-Ray Bursts (GRBs). This
was followed by a detection of a strong unusual radio flare from
the same source (Kulkarni et al. 1999, Galama et al. 1999). One of
the most impressive facts concerning the strong optical flash of
GRB 990123 was that such a strong prompt optical emission was
predicted, just a few months earlier to arise from the reverse
shock during the early afterglow (Sari, \& Piran 1999a). Later
this interpretation was questioned and other models for prompt
optical emission were proposed (e.g. Beloborodov, 2002). We return
to this issue here.

In the popular fireball model (Piran 1999) for Gamma-Ray Bursts
(GRBs), the afterglow results from a blast-wave that propagates
into the circum-burst medium. This blast-wave is originated by the
 energy dissipation  of the relativistic ejecta to the circum-burst
medium. In this model the early afterglow is produced during this
dissipation, and thus can be used to study the nature of the
ejecta. In case that the ejecta is Baryonic, a reverse shock is
produced during this dissipation. This reverse shock is predicted
to produce strong optical flash and radio flare (Sari \& Piran
1999b,c).

Recently we have presented new theoretical analysis of the reverse
shock emission (Nakar \& Piran 2004; hereafter NP04). This
analysis includes a new test of reverse shock emission and
diagnostic tools for the physical parameters of the original
ejecta. We use these new tools, here, to re-investigate the early
afterglow of GRB 990123. Note that numerous authors analyzed the
early afterglow of GRB 990123 (Sari \& Piran 1999b; M\'esz\'aros
\& Rees 1999; Kulkarni et al. 1999; Galama et al. 1999; Wang, Dai
\& Lu 2000; Panaitescu \& Kumar 2001; Fan et al. 2002; Soderberg
\& Ramirez-Ruiz 2002).  We don't attempt to review here these
results and compare them as the scope of such a review is too long
for a short letter.

The reverse shock optical emission has a typical and
characteristic $t^{-2}$ decay  that was observed in the light
curve of GRB 990123  (Fenimore et al. 1999; Sari \& Piran
1999b,c). We find additional conclusive evidence that the optical
flash and the radio flare of GRB 990123 indeed resulted from the
emission of a reverse shock produced by an interaction of the
ejecta with and ISM. These results suggest first that at least in
this burst the reverse shock emission, and not another mechanism
such as a pair enriched forward shock (due to interaction of the
prompt $\gamma$-rays with the circum burst medium; Thompson \&
Madau 2000, Beloborodov, 2002), dominated the early afterglow. As
the reverse shock requires, a baryonic component in the ejecta it
also implies that a significant fraction of the energy of the
relativistic ejecta was carried by baryons.

The sparse observations before the peak of the optical flash (only
one measurement before the peak at each wavelength) poses
difficulties in analyzing the properties of the ejecta, since this
analysis depends on the rising power-law index of the optical
light curve\footnote{Since there is only a single observation
before the peak, the rising phase may also be very irregular with
no power-law behavior. In this case the analysis of the ejecta
properties presented in NP04 does not apply, as it imply that the
relativistic ejecta hydrodynamical profile is highly irregular.
However, having no detailed observations of this phase we use here
the simplest light curve that fits the observations - a broken
power-law. Note that the tests, which show that the emission is a
reverse shock emission, are independent of the shape of the rising
light curve, and can be carried also if it is irregular.}. Still,
we are able to determine that the reverse shock was mildly
relativistic and that the initial Lorentz factor is most likely
smaller than $\sim 600 (E_{54}/n)^{1/8}$, where $E_{54}$ is the
(isotropic equivalent) kinetic energy in the ejecta after the
prompt emission phase in units of $10^{54}$ergs and $n$ is the
protons density of the circum-burst interstellar medium in c.g.s.
The width of the ejecta is found to be, as expected, roughly equal
to the burst duration (multiplied by the light speed).  Because of
the sparse data available we obtain only upper limits to the
initial Lorentz factor and the shell's width and only lower limits
to the microscopic equipartition parameters and the external
density. We expect that in the future detailed optical light
curves, provided by Swift and rapid follow up observations, would
enable an accurate determination of the initial physical
parameters of the relativistic wind by using the method we apply
here.

For completeness  we begin (\S \ref{sec Theory}) with a short
summary of the theoretical model of the reverse shock emission
presented in NP04. We summarize, in \S \ref{sec Observations}, the
early afterglow observations of GRB 990123. The analysis of the
optical light curve and the resulting constraints on the physical
parameters of the relativistic wind is presented in \S \ref{sec
analysis}. In \S \ref{sec test} we test whether the observations
fit with a reverse shock emission. The results,  their
implications to GRB modelling and future prospects are summarized
in \S \ref{sec Conclusions}.

\section{The reverse shock emission - Theory}\label{sec Theory}

We  begin with a brief summary of the main results of NP04. These
results give the theoretical predictions of the optical and the
radio emission of a baryonic reverse shock in an ISM environment.
NP04 suggest to model the early optical light curve  (similarity
to late afterglow parametrization; Beuermann et al. 1999) as a
broken power-law with five parameters:
\begin{equation}\label{EQ FnurParam}
    F_{\nu,opt}^r(t)=F_0^r \left( \frac {1}{2} \left(  \frac{t}{t_0} \right) ^{-s\alpha_1}
    +  \frac {1}{2} \left( \frac{t}{t_0} \right) ^{-s\alpha_2} \right)
    ^{-\frac{1}{s}} ,
\end{equation}
where $\alpha_1>0$ and $\alpha_2<0$ are the asymptotic power-law
indices, $F_{\nu,opt}^r(t_0)=F_0^r$ is the peak flux and the
parameter $s$ determines the sharpness of the peak  (a large $s$
corresponds to a sharp peak). The physical parameters which
determine the emission of the reverse shock are: (1) The
(isotropic equivalent) energy $E$, the width $\Delta$ and the
initial Lorentz factor $\Gamma_o$ of the ejected wind. (2) The
circum burst density, $n$, that we consider as a constant
interstellar medium (ISM). (3) The microscopic parameters in the
emitting region, namely the energy equipartition parameters
$\epsilon_e$ and $\epsilon_B$ that describe the ratio of the
electrons and Magnetic field energy to the total internal energy,
and $p$, the power-law index of the electrons' energy
distribution.

The strength of the reverse shock is determined by the
dimensionless parameter $\xi$ (Sari \& Piran 1995):
\begin{equation}\label{EQ xi}
    \xi\equiv\frac{l^{1/2}}{\Delta^{1/2}\Gamma_o^{4/3}} ,
\end{equation}
where $l \equiv (3E/(4\pi nm_pc^2))^{1/3}$ ($m_pc^2$ is the proton
rest mass energy). According to this definition:
\begin{equation}\label{EQ gamma0}
    \Gamma_o=188\xi^{-3/4}\Delta_{12}^{-3/8}(E_{52}/n)^{1/8} .
\end{equation}
The peak of the optical emission is observed at
\begin{equation}\label{EQ t0}
    t_0=\frac{\Delta}{c}(1+0.7\xi^{3/2})(1+z) ,
\end{equation}
when the reverse shock finishes crossing the ejecta. $c$ is the
speed of light and $z$ is the redshift.  The structure of the
optical and the radio reverse shock light curves depend on the
relative values of the three reverse shock break frequencies at
$t_0$: $\nu_a^r$, the self-absorbtion frequency, $\nu_m^r$, the
synchrotron frequency and $\nu_c^r$ the cooling frequency. In
principal numerous patterns are possible. However, over a wide
range of the parameter space (and as we show below, in
particularly   for GRB 990123) these frequencies satisfy
$\nu_{radio}<\nu_m^r(t_0)<\nu_a^r(t_0)<\nu_{opt}<\nu_c^r(t_0)$,
where $\nu_{radio}$ and $\nu_{opt}$ are the observed radio and
optical frequencies respectively. We consider this frequencies
sequence as the {\it generic case}. In this case the peak optical
flux is:
\begin{equation}\label{EQ F0r}
    F_0^r=0.1{\rm mJy}(1+z)^{-\frac{4+p}{8}}1.5^{2.5-p}
    \left(\frac{3(p-2)}{p-1}\right)^{p-1}\epsilon_{e-1}^{p-1}
    \epsilon_{B-2}^{\frac{p+1}{4}}n^{\frac{p+2}{8}}E_{52}^{1+\frac{p}{8}}
    t_{0,2}^{-\frac{3p}{8}}D_{28}^{-2}A_{F,0}^r(\xi),
\end{equation}
where throughout the paper we denote by $Q_{x}$ the value of the
quantity $Q$ in units of $10^{x}$ (c.g.s), and $D$ is the proper
distance. The function $A_{F,0}^r(\xi)$ is approximated in the
range of $0.1<\xi<2.5$ by:
\begin{equation}\label{EQ AF0rApprox}
    A_{F,0}^r(\xi) \approx 180 \xi^{0.65} \left(6 \cdot 10^{-4}
    \xi^{-2.6}\right)^\frac{p-1}{2}.
\end{equation}
$A_{F,0}^r(\xi)$  values out of this range can be found in NP04.

The flux before $t_0$ depends strongly on $\xi$, and thus the
rising slope, $\alpha_1$ can be used to determine $\xi$ using (in
the generic case):
\begin{equation}\label{EQ alpha1}
    \alpha_1 \approx 1.2(0.5+\frac{p}{2}(\xi-0.07\xi^2)).
\end{equation}
The reserve shock decay index is, however, a constant:
\begin{equation}\label{EQ alpha1}
    \alpha_2 \approx -2,
\end{equation}
and its main use is  to identify the reverse shock emission.
Finally, the sharpness parameter, $s$, depends mainly on the
profile of the ejecta. A sharp break with $s \gtrsim 3$  is
expected when the ejecta is rather homogenous, while a gradual
break with a lower value of $s \sim 1$ is expected if the spread
of the initial Lorentz factor within the shell is large.

In contrast to the optical emission, the radio continues to rise
at $t>t_0$ and it peaks at a later time, $t_*$, when
$\nu_{radio}=\nu_a^r$. Over a wide range of $\xi$ values (when the
shock is not ultra relativistic) $\nu_{radio}<\nu_m^r(t_0) <
\nu_a^r(t_0) \approx 10^{12-13}$Hz. In this case the radio
emission is expected to rise first as $\sim t^{0.5}$ until
$\nu_{radio}=\nu_m^r$ and then as $\sim t^{1.25}$ until $t_*$. At
$t>t_*$ it is expected to decay, similarly to the optical
emission, as $\sim t^{-2}$. This behavior provides a second and
independent test for the reverse shock emission:
\begin{equation}\label{EQ radioTest}
\frac{F_*}{F_0}\left(\frac{t_*}{t_0}\right)^{\frac{p-1}{2}+1.3}=\left(\frac{\nu_{opt}}{\nu_{radio}}\right)^\frac{p-1}{2}
\sim 1000,
\end{equation}
where this value can be larger or smaller by a factor of $\sim 3$
(for a given $p$) due to uncertainty in the hydrodynamics (see
Kobayashi \& Sari 2000). It provides also an independent
measurement of $\nu_a^r(t_0)$:
\begin{equation}\label{EQ nuat0}
\nu_a^r(t_0)\approx \frac{t_*}{t_0} \nu_{radio}
\end{equation}
Detailed radio observations at $t<t_*$ that identify the break
during the rising phase, when $\nu_m^r=\nu_{radio}$, would enable
determination of $\nu_m^r(t_0)$ as well.

\section{Observations}\label{sec Observations}
At $z=1.6$ GRB 990123 is one of the brightest $\gamma$-rays bursts
observed so far (Briggs et al. 1999). The main event of this burst
is composed of two very energetic and hard pulses which last
together $\sim 25$sec. These two pulses contain almost all the
emitted energy during the burst. A much softer and less energetic
emission ($<120$keV) is observed $\sim 20$sec before and up to
$\sim 50$sec after this event. The total isotropic equivalent
energy emitted in $\gamma$-rays during the burst is $\sim
10^{54}$ergs.  The timing of the main $\gamma$-rays event differ
between different detectors. While BATSE triggered on the first
soft emission, $\sim 20$sec before the main event, BeppoSAX
triggered $18$sec later than BATSE on the beginning of the main
event.

GRB 990123 is also the only burst where optical emission was
observed during the prompt emission (Akerlof et al. 1999). The
first detection  of $11.7$mag is recorded $\sim 7 \rm sec$ after
BeppoSAX trigger.  The optical emission peaks at the second
snapshot that took place $\sim 32$sec after BeppoSAX trigger with
$8.9$mag ($\approx 0.8$Jy)! After the peak, a fast decay is
observed with a power law slope of $\approx -2$ which becomes
shallower at later times.

The early radio observations of GRB 990123 show also an unusual
flare (Kulkarni et al. 1999). The first observation after
$0.25$days shows a 8.46GHz intensity of $62\pm 32\mu$Jy. The flux
in this band rises to a peak flux of $260\pm 32\mu$Jy after
$1.25$days and then decline rapidly. Observations at other
wavelengths from this epoch ($\sim 1$day) (Galama et al. 1999)
show a flux of $118 \pm 40\mu$Jy at $4.88$GHz and upper limits of
several hundreds $\mu$Jy in $15$, $86$ and $222$GHz.

\section{Constraining the burst parameters} \label{sec analysis}

Already in 1999 Fenimore et al. (1999) pointed out that the
initial decay index of the optical flash $\alpha_2 \approx -2$
agrees with the predictions of  a reverse shock emission  (Sari \&
Piran 1999a,b,c). At later times the decay becomes more moderate
as expected from the forward shock contribution.  Motivated by
this observation, we first use the optical observations to
constrain the physical parameters of the burst assuming that the
flash arises from  a reverse shock emission. Later, we carry the
additional tests of reverse shock emission, described in NP04.

When modelling the early afterglow, it is important to find the
time when the main part of the relativistic ejecta is emitted from
the source. It is not necessarily the trigger time. In GRB 990123
the radiated energy, and therefore most likely the energy ejected
from the source, is clearly dominated by the main two pulses.
Kobayashi, Sari \& Piran 1997 (see also Nakar \& Piran 2002), have
shown that in internal shocks the observed time of the
$\gamma$-rays pulses reflects the emission time of the wind from
the source. Therefore we estimate that the main part of the
relativistic wind is beginning to be ejected by the source at the
time that the first dominant pulse starts rising - the BeppoSax
triggering time ($18$sec after the BATSE trigger). This is the
point where we set the observer clock to $t=0$. According to this
setting the duration of the burst is $\sim 25$sec and the peak of
the optical emission is observed at $t=32$sec.

The observations of the optical flash of GRB 990123 (Akerlof et
al. 1999) include only one data point before the peak. Since there
are two observations after the peak which show similar decay
($t=57$sec and $t=142$sec), $\alpha_2 \approx -2$, a broken power
law fit is possible only if the peak is between the first and the
second observations. Thus, we have only lower limits to
$F_0>0.8$Jy, $\alpha_1>2$, an upper limit to $t_0<32$sec  and $s$
can take any value (see Fig. 1). As we have only a lower limit on
$\alpha_1$, Eq. (\ref{EQ alpha1}) yields only a lower limit on
$\xi$ (we use $p=2.3$ thorough out.):
\begin{equation}\label{EQ xi990123}
    \xi \gtrsim 1.
\end{equation}
 Eqs. (\ref{EQ t0} \& \ref{EQ gamma0}) result in the upper
 limits:
\begin{equation}\label{EQ D990123}
    \Delta \lesssim 2.5 \cdot 10^{11} \rm cm,
\end{equation}
and
\begin{equation}\label{EQ Gamma990123}
    \Gamma_o  \lesssim 600 \left(\frac{E_{54}}{n}\right)^{1/8} .
\end{equation}
It is reassuring to find  that the width of the shell obtained for
$\xi \approx 1$ divided by the speed of light is similar to the
duration of the GRB ($\sim 10$sec in the burst's frame).

Next we use Eq. (\ref{EQ F0r}) to constrain the microscopic
equipartition parameters, $\epsilon_e$ and $\epsilon_B$, and the
external density, $n$. We consider the total energy emitted in
$\gamma$-rays, $\sim 10^{54}$ergs,  as a reasonable estimate of
the remaining energy in the ejecta, $E$. Therefore, we do not
attempt to estimate it from the optical flash observations.
Instead we express the dependance of the resulting constraints on
the value of $E$. Taking $\xi>1$ and requiring $F_0>0.8$Jy leads,
using  Eq. (\ref{EQ F0r}), to:
\begin{equation}\label{EQ constraint}
    \epsilon_{e-1}^{p-1}
    \epsilon_{B-2}^{\frac{p+1}{4}}n^{\frac{p+2}{8}} \gtrsim 15E_{54}^{-(1+p/8)}
\end{equation}
Another constraint follows the requirement that
$\nu_{opt}<\nu_c^r(t_0)$ (otherwise the optical light curve would
not show the generic behavior of $\alpha_2 \approx 2$) \footnote{
The requirement that $\nu_{radio}<\nu_m^r(t_0)<\nu_{opt}$ is
satisfied for any reasonable value of the parameters and therefore
cannot be used as a constraint. The values of $\nu_m^r(t_0)$ and
$\nu_c^r(t_0)$ can be found in NP04.}:
\begin{equation}\label{EQ nuc_constraint}
    n \epsilon_{B-2}^{3/2} \lesssim 20E_{54}^{-1/2}.
\end{equation}
These two constraints (together with the trivial one, $\epsilon_e+
\epsilon_B<1$) result in the following limits:
\begin{equation}\label{EQ ee constraint}
    \epsilon_e \gtrsim 0.1 E_{54}^{-0.75},
\end{equation}
\begin{equation}\label{EQ eb constraint}
    5\cdot 10^{-3} E_{54}^{-1.6} n^{-2/3} \lesssim \epsilon_B \lesssim 0.1 E_{54}^{-1/3} n^{-2/3},
\end{equation}
\begin{equation}\label{EQ ee constraint}
    n \gtrsim 5 \cdot 10^{-3} E_{54}^{-2.4} \rm cm^{-3}
\end{equation}

\section{Was the optical flash a result of a reverse shock
emission?}\label{sec test}

We have already mentioned that the optical flash of GRB 990123
passes the first test of a reverse shock emission: $\alpha_2
\approx -2$ (Fenimore et al. 1999, Sari \& Piran 1999a,b,c). NP04
have shown that in the generic case
($\nu_{radio}<\nu_m^r<\nu_a^r<\nu_{opt}<\nu_c$ at $t_0$) the
reverse shock emission results also in a tight relation between
the radio and the optical emission, Eq. (\ref{EQ radioTest}). For
the observed  frequencies of GRB 990123 ($\nu_{opt}=5\cdot
10^{14}$Hz and $\nu_{radio}=8.6$GHz) and $p=2.3$:
\begin{equation}\label{EQ radioobs1}
\left(\frac{\nu_{opt}}{\nu_{radio}}\right)^\frac{p-1}{2} =1300,
\end{equation}
The observations of GRB 990123 show a remarkable agreement with
this prediction (taking the times and the fluxes of the flash and
the flare as the times and the fluxes of the peak observation,
$t_0=32$sec, $F_0=0.8$Jy, $t_*=1.25$day and $F_*=260\mu$Jy):
\begin{equation}\label{EQ radioObs2}
    \frac{F_*}{F_0}\left(\frac{t_*}{t_0}\right)^{\frac{p-1}{2}+1.3}
    \approx 800-4000,
\end{equation}
where the result include the uncertainty in the hydrodynamics.
NP04 show also that if the above test is passed, then the self
absorption frequency at $t_0$, $\nu_a^r(t0)$, can be estimated
from the the ratio between $t_*$ and $t_0$:
\begin{equation}\label{EQ nuaObs}
    \nu_a^r(t_0) \approx \frac{t_*}{t_0}\nu_{radio}=3 \cdot
    10^{13}Hz.
\end{equation}
This frequency is determined by the physical conditions at $t_0$.
As a consistency check we compare this result to the possible
values of $\nu_a^r(t_0)$ obtained by the limits  on the physical
parameters determined from the analysis of the optical emission.
We consider the case of $\xi=1$ and p=2.3 for which:
\begin{equation}
\nu_a^r(t_0)=6 \cdot 10^{12}Hz \left[(1+z)^{-\frac{p+6}{8}}
\epsilon_{e,-1}^{\;p-1}\epsilon_{B,-2}^\frac{p+2}{4} (n
E_{54})^\frac{p+6}{8}t_{0,2}^{-\frac{3p+10}{8}}
\right]^\frac{2}{p+4}.
\end{equation}
Within the parameters space of $\epsilon_e$,$\epsilon_B$ and $n$
which satisfy Eqs. (\ref{EQ constraint} \& \ref{EQ
nuc_constraint}) and taking $E_{54}=1$ we obtain $1.5 \cdot
10^{13}Hz<\nu_a^r(t_0)<3.5 \cdot 10^{13} n^{1/6}Hz$! Although the
dependance of this range on the value of $E$ is not trivial, it is
weak and the value $\nu_a^r(t_0)=3 \cdot 10^{13}Hz$ is consistent
with the allowed range for any reasonable value of $E$,
$0.5<E_{54}<10$. This constraint on the value of $\nu_a^r(t_0)$ is
independent of the value found in Eq. (\ref{EQ nuaObs}). Once more
the agreement between the two estimates is remarkable.

So far we focused on the radio observations at 8.46Ghz. The  lower
radio fluxes both above and below 8.46GHz, at $t_*$, are
consistent with reverse shock emission, which predicts that the
peak frequency at $t_*$ is the observed $\nu_{radio}$ (8.46Ghz in
this case). Finally we note that although radio observations from
$t<t_*$ can theoretically be used to farther constrain the
conditions at $t_0$, the large errors of the single radio
observation from this epoch is insufficient to do so.

\section{Conclusions}
\label{sec Conclusions}

 We have analyzed the early afterglow of GRB
990123 according to the theoretical predictions of the reverse
shock emission presented in NP04. We find that GRB 990123 shows
the clear signature of a reverse shock emission in two independent
and robust tests. Apart from these two tests we carry a
consistency check between the radio emission and our analysis of
the optical emission which is passed successfully. This
consistency check gives us further confidence both in the analysis
and in the fact that the optical flash and the radio flare of GRB
990123 are a generic case of a reverse shock emission.

The consistency with the reverse shock model further suggests, but
does not proof of course, that the optical flash of GRB 990123 did
not arise from a pair loaded forward shock (Beloborodov, 2002). In
turn this indicates that the ejecta of GRB 990123 had (at least at
this stage) a significant baryonic component, as otherwise, for
example in a pure Poynting flux flow, a reverse shock is not
expected.

Our analysis of the optical emission shows that the reverse shock
was mildly relativistic and that the initial Lorentz factor is
less than $600(E_{54}/n)^{1/8}$. We find that the width of the
initial shell is similar to the duration of the burst or smaller.
We constrain also the equipartition parameters and find that
$\epsilon_e \gtrsim 0.1 E_{54}^{-0.75}$ while $5\cdot 10^{-3}
E_{54}^{-1.6} n^{-2/3} \lesssim \epsilon_B \lesssim 0.1
E_{54}^{-1/3} n^{-2/3}$. The external density is limited by $n
\gtrsim 5 \cdot 10^{-3} E_{54}^{-2.4}\rm cm^{-3}$.

The determination of the initial conditions of GRB 990123 and the
constraints of the microscopic parameters and the external density
is very limited due to the sparse optical and radio observations.
Swifts and its follow up observations would hopefully provide
during the next few years detailed light curves. These would
enable a more detailed analysis and a much better determination of
the parameters of the burst's relativistic outflow.

We thank Robert Mochkovitch, Frederic Daigne  and  Elena Rossi for
helpful discussions. The research was supported by the US-Israel
BSF and by EU-RTN: GRBs - Enigma and a Tool. EN is supported by
the Horowitz foundation and by a Dan David Prize Scholarship 2003.

\begin{figure}[h]
\begin{center}
\includegraphics[width=14cm,height=8cm]{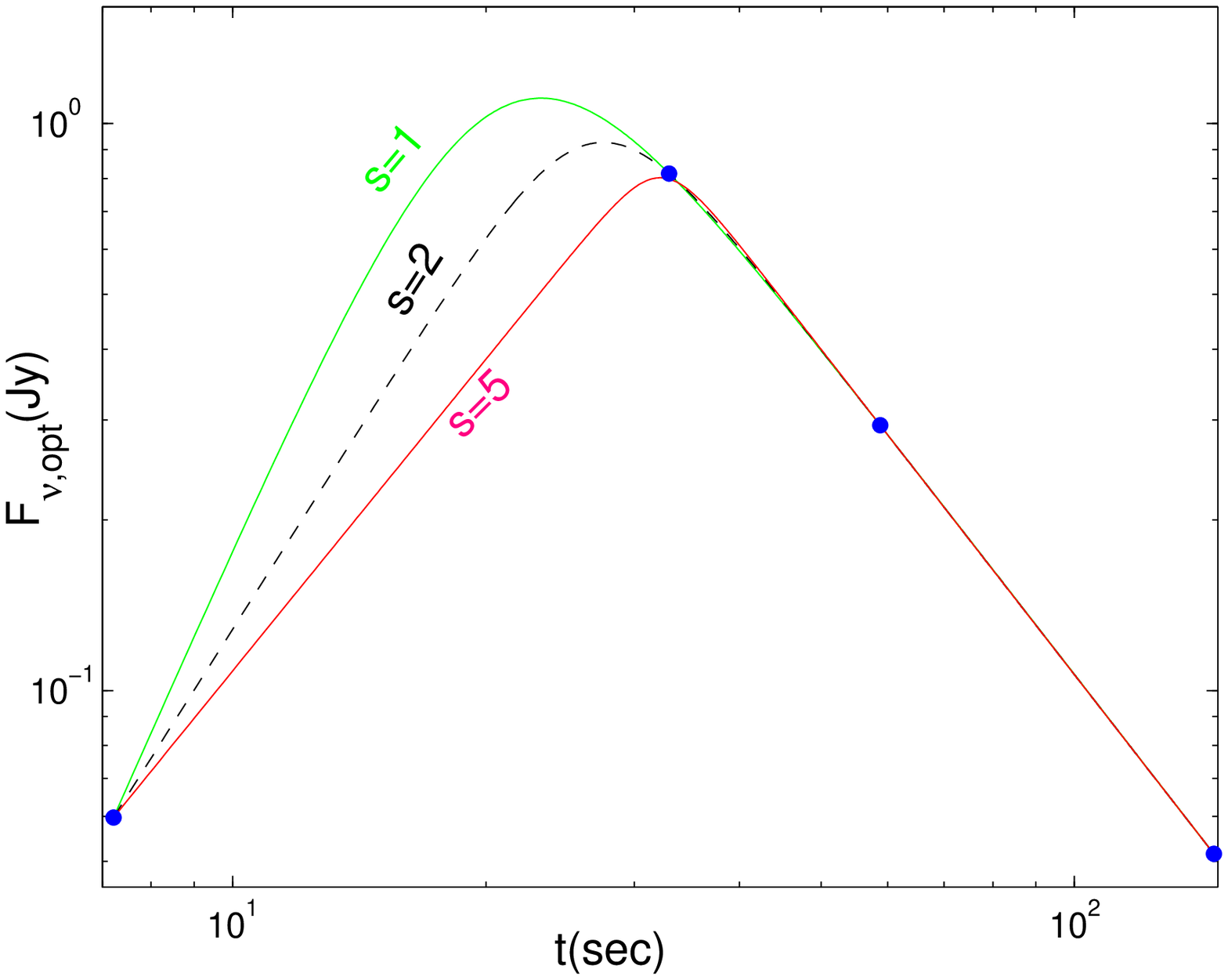}
\caption{Possible light curves (with the value of $s$ beside each
light curve) fitted according to Eq. (\ref{EQ FnurParam}) to the
optical observations ({\it full dots}) of GRB 990123.}
\label{plotone}
\end{center}
\end{figure}

\end{document}